\pdfoutput=1
\documentclass[cits]{JINST}

\title{Measurements of the \v{C}erenkov light
  emitted by a TeO$_2$ crystal}

\author{F.~Bellini$^{a,b}$,
  N.~Casali$^{b,c}$,
  I.~Dafinei$^b$,
  M.~Marafini$^d$,
  S.~Morganti$^b$,
  F.~Orio$^b$,
  D.~Pinci$^b$\thanks{Corresponding author.},
  M.~Vignati$^b$ and
  C.~Voena$^b$\\
\llap{$^a$}Sapienza Universit\`a di Roma, 
P.le A. Moro 2, Roma, Italy\\
\llap{$^b$}Istituto Nazionale di Fisica Nucleare, Sezione di Roma, 
P.le A. Moro 2, Roma, Italy\\
\llap{$^c$}Universit\'a degli Studi dell'Aquila, Italy\\
\llap{$^d$}Museo Storico della Fisica 
e Centro Studi e Ricerche "Enrico Fermi",
Piazza del Viminale 1, Roma, Italy\\

E-mail: \email{davide.pinci@roma1.infn.it}}

\abstract{Bolometers have proven to be good instruments to search for rare processes because of their 
excellent energy resolution and their extremely low intrinsic background.
In this kind of detectors, the capability of discriminating alpha particles from electrons 
represents an important aspect for the background reduction.
One possibility for obtaining such a discrimination is provided by the detection of the \v{C}erenkov 
light which, at the low energies of the natural radioactivity, is only emitted by electrons.

In this paper, the results of the analysis of the light emitted by a TeO$_2$ crystal at 
room temperature when transversed by a cosmic ray are reported.
Light is promptly emitted after the particle crossing 
and a clear evidence of its directionality is also found.

These results represent a strong indication that \v Cerenkov light is the main, if not even the only, 
component of the light signal in a TeO$_2$ crystal. 
They open the possibility to make large improvements 
in the performance of experiments based on this kind of materials.}


\begin{document}

\section{Introduction}

Tellurium dioxide (TeO$_2$) crystals have proven to be superb bolometers for the 
search of neutrinoless double beta decay\cite{bib:cuore,bib:cuoricino}.
They are able to measure energies in the MeV region with a resolution of
the order of few keV.
One of the main sources of background in these searches is 
represented by the $\alpha$ particles emitted by natural radioactivity.
As predicted in~\cite{TabarellideFatis:2009zz} and demonstrated in~\cite{bib:cherenkov},
the observation of light emitted by electrons in a TeO$_2$ bolometer
can provide a powerful tool to disentangle 
$\alpha$ from $\beta/\gamma$ radiation. 
According to these results, the detected light was compatible with the \v Cerenkov emission, 
even though the scintillation hypothesis could not be discarded.
The aim of the experiment presented in this paper is the assessment and the measurement 
of the \v{C}erenkov contribution in the light yield of a TeO$_2$ crystal.
In order to distinguish it from a possible scintillation emission,
the differences between these two processes can be exploited.
The scintillation light is isotropically emitted and usually shows 
a time development with an exponential decay typical of the material.
\v{C}erenkov light is instead promptly emitted when a charged particle 
crosses a material with a velocity larger than the speed of 
light in that material.
Moreover, \v{C}erenkov photons are emitted in a cone with an opening angle 
$\theta_c\;=\;$arccos(1/($\beta n$)) with respect to the particle direction.
As it was already demonstrated \cite{bib:wig0}-\cite{bib:wig4},
the study of the signal shape and of the directionality of the light yield
represents an useful tool to disentangle these two components.

\section{Experimental set-up}

\begin{figure}[h]
\begin{centering}
\includegraphics[width=10cm]{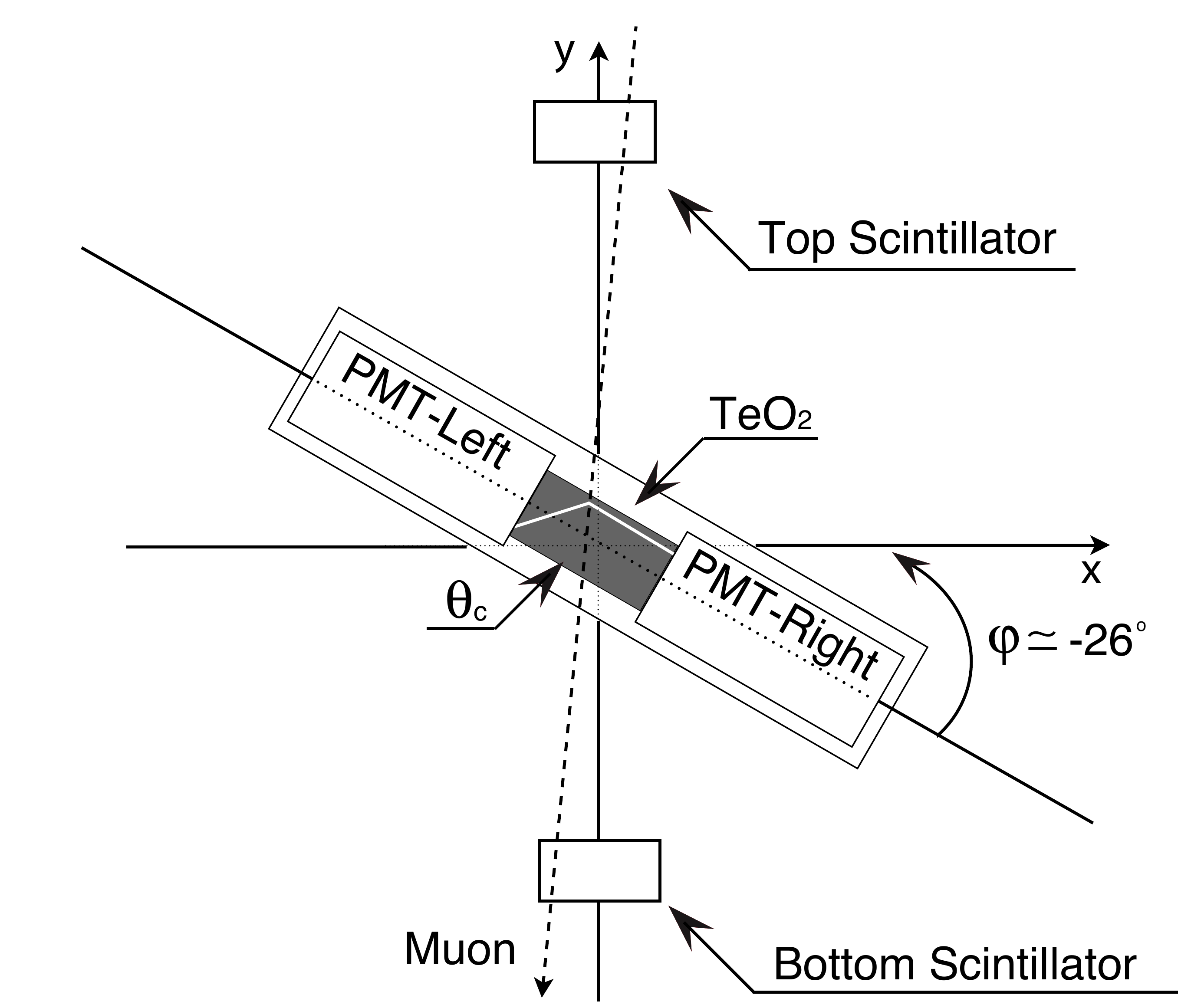}
\caption{Experimental setup. See text for details.}
\label{fig:setup}
\end{centering}
\end{figure}

To perform the measurements of the light produced by a TeO$_2$ crystal,
the set-up shown in Fig.~\ref{fig:setup} was built. 
A $5\times2.5\times2.5$ cm$^3$ crystal placed inside a black box 
was read-out on the two small opposite faces with two photo-multiplier tubes (PMTs)
XP2970\footnote{10-stages, UV-Sensitive, 29 mm diameter. 
For more information http://www.photonis.com/en/ism.php}. 
These tubes were chosen for their extended sensitivity in the UV region
where the production of \v{C}erenkov photons is expected to be large.
They were operated at a voltage of 1200~V, with an expected gain of about 10$^7$.
Their analog signals were sent to a CAEN V1371 8-bit digitizer working with
1 GS/s sampling rate.

The box was free to rotate in the $XY$ plane giving the possibility of 
changing the angle $\varphi$ between the longest crystal axis
and the horizontal direction in the range $\pm \;40^{\circ}$.
The maximum \v Cerenkov light transmission to a PMT is expected 
with the crystal parallel to the \v Cerenkov photon direction.
Since the refractive index in the band of the detected light 
is about 2.27,
for an angle $\varphi_m~=~90^{\circ}~-~\theta_c~=~26^{\circ}$,
PMT-Left is expected to see the maximum amount of \v Cerenkov light
which, instead, reaches PMT-Right for $\varphi~=~-\;26^{\circ}$.
In order to select vertical muons in cosmic rays,
the trigger signal to the data acquisition,
was provided by the coincidence of 
two 2~cm thick, 4~$\times$~7~cm$^2$ scintillator fingers placed above and
below the crystal.
The distance between the scintillators was of about 50~cm 
and the trigger rate was about 0.1 Hz.

\section{Behavior of the detected light}

The light exiting from a face of the crystal can
be separated into two components:
\begin{itemize}
\item {\bf A}: a part that is independent from the angle between the muon and the crystal. 
This light can be scintillation light 
or \v Cerenkov light diffused by the internal reflections on the crystal faces
loosing its initial directionality.
\item {\bf B($\varphi$)}: a component produced with a directionality and for which the probability of
exiting from a face of the crystal is a function of the angle $\varphi$. This component is expected
to be entirely due to the \v Cerenkov light.
\end{itemize}
The total light exiting on the two lateral faces of the crystal will result:

\vbox{\begin{eqnarray}
\bar{L}(\varphi)&=&\frac{\alpha}{\mbox{cos} \varphi}\left(A_L+B_L(\varphi)\right) \\
\bar{R}(\varphi)&=&\frac{\beta} {\mbox{cos} \varphi}\left(A_R+B_R(\varphi)\right)
\end{eqnarray}}

with $\alpha$ and $\beta$ being two parameters that take into account
possible non-equalizations of the PMT responses while 1/cos$\varphi$ 
is proportional to the path length of the muon within the crystal.
Because of symmetry reasons, one expects:
\begin{eqnarray}
A_L&=&A_R=A \\
B_L(\varphi)&=&B_R(-\varphi)=B(\varphi)
\end{eqnarray}
For $\varphi=0$ it follows:
\begin{eqnarray}
\bar{L}(0)={\alpha}\left(A+B(0)\right) = \alpha k\\
\bar{R}(0)={\beta} \left(A+B(0)\right) = \beta k
\end{eqnarray}
Defining L($\varphi$) and R($\varphi$) as the responses equalized at $\varphi=0$, it follows:
\begin{eqnarray}
L(\varphi)&=&\frac{\bar{L}(\varphi)\mbox{cos} \varphi}{\bar{L}(0)}=
\frac{1}{k}\left(A+B(\varphi)\right) \\
R(\varphi)&=&\frac{\bar{R}(\varphi)\mbox{cos} \varphi}{\bar{R}(0)}=
\frac{1}{k}\left(A+B(-\varphi)\right).
\end{eqnarray}
%From the symmetry, it results that:
%\begin{equation}
%L(\varphi)=R(-\varphi).
%\end{equation}

\section{Waveform analysis}

\begin{figure}
\begin{centering}
\includegraphics[width=10cm]{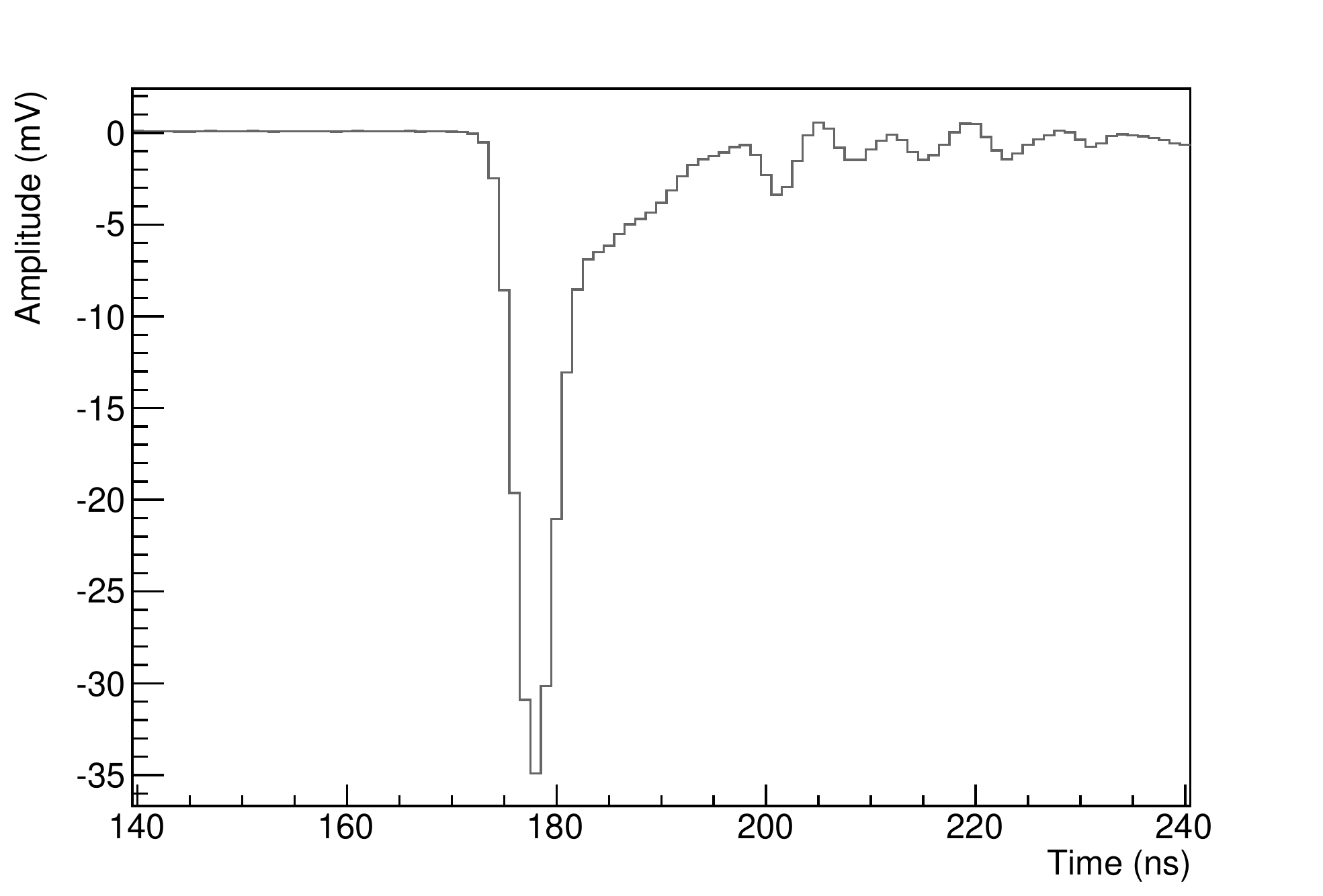} 
\caption{Average signal shape of PMT-Right.}
\label{fig:average}
\end{centering}
\end{figure}

The waveforms of the signals provided by the two PMTs are acquired and 
off-line analyzed.
The average waveform of PMT-Right obtained for a thousand muon events is
shown in Fig.~\ref{fig:average}.
The signals show a rise time and a decay time of the order of few nanoseconds. 
This very fast behavior is a first indication that an important component of the light is due to
\v Cerenkov emission.

\begin{figure}
\begin{centering}
\includegraphics[width=10cm]{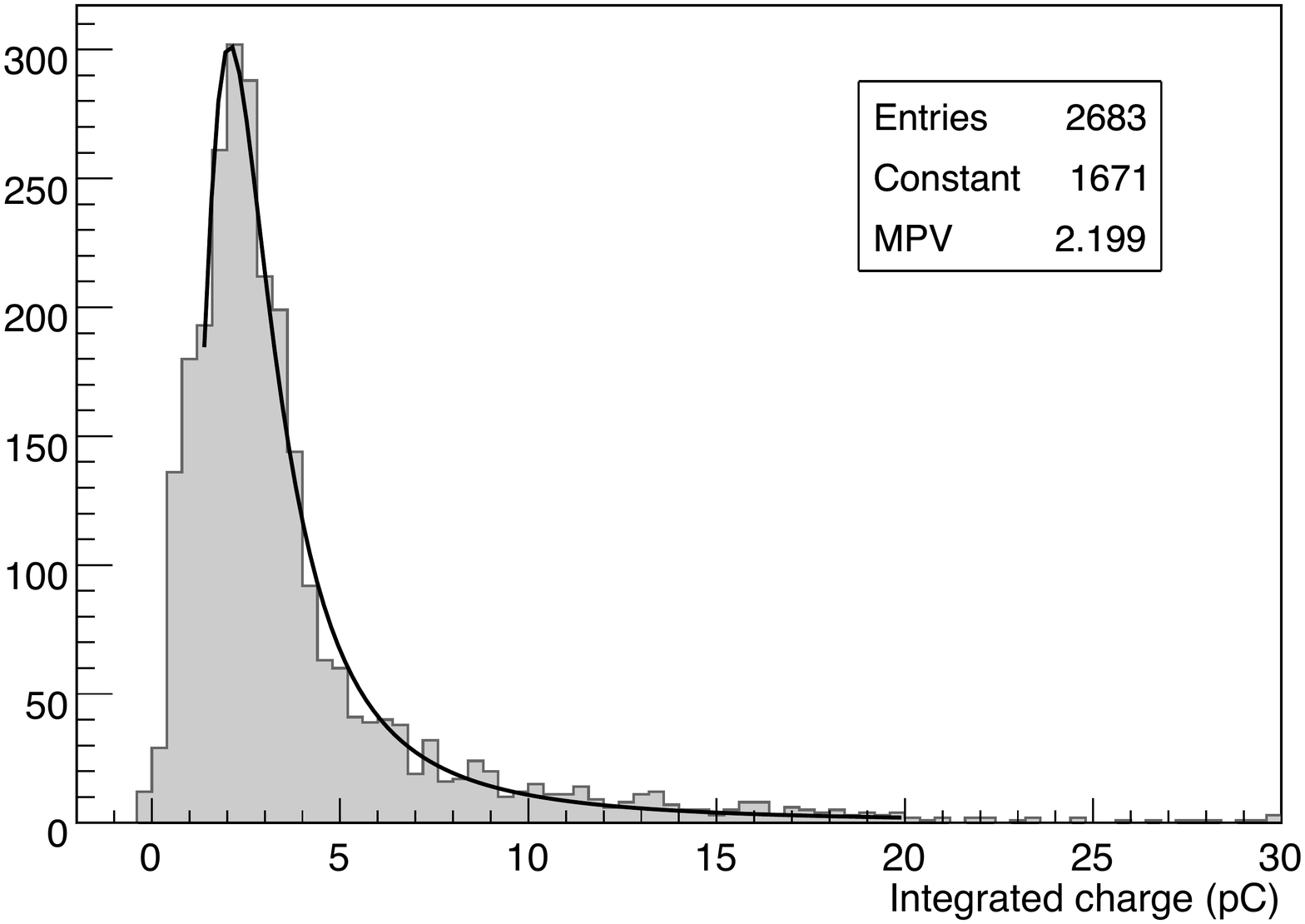}
\caption{Example of the charge spectrum.}
\label{fig:spectrum}
\end{centering}
\end{figure}

In order to evaluate the charge produced by the PMT, the waveforms are integrated, event by event,
in a 15 ns wide time window around the maximum signal amplitude. 
An example of a charge spectrum obtained by the PMT-Right
is shown in Fig.~\ref{fig:spectrum}.
The fit of the charge spectrum with a Landau function returns
the average charge and thus an evaluation of the light yield.
The effect of the electronics noise is computed by integrating the same 
waveforms in a 15~ns time interval before the signal pulse. 
The width of the pedestals resulted to be about the 4\% of the FWHM of the distribution.
Therefore, the effect of the electronics noise is negligible.

\section{Results from the angular scan}

\begin{figure}[h]
\begin{centering}
\includegraphics[width=10cm]{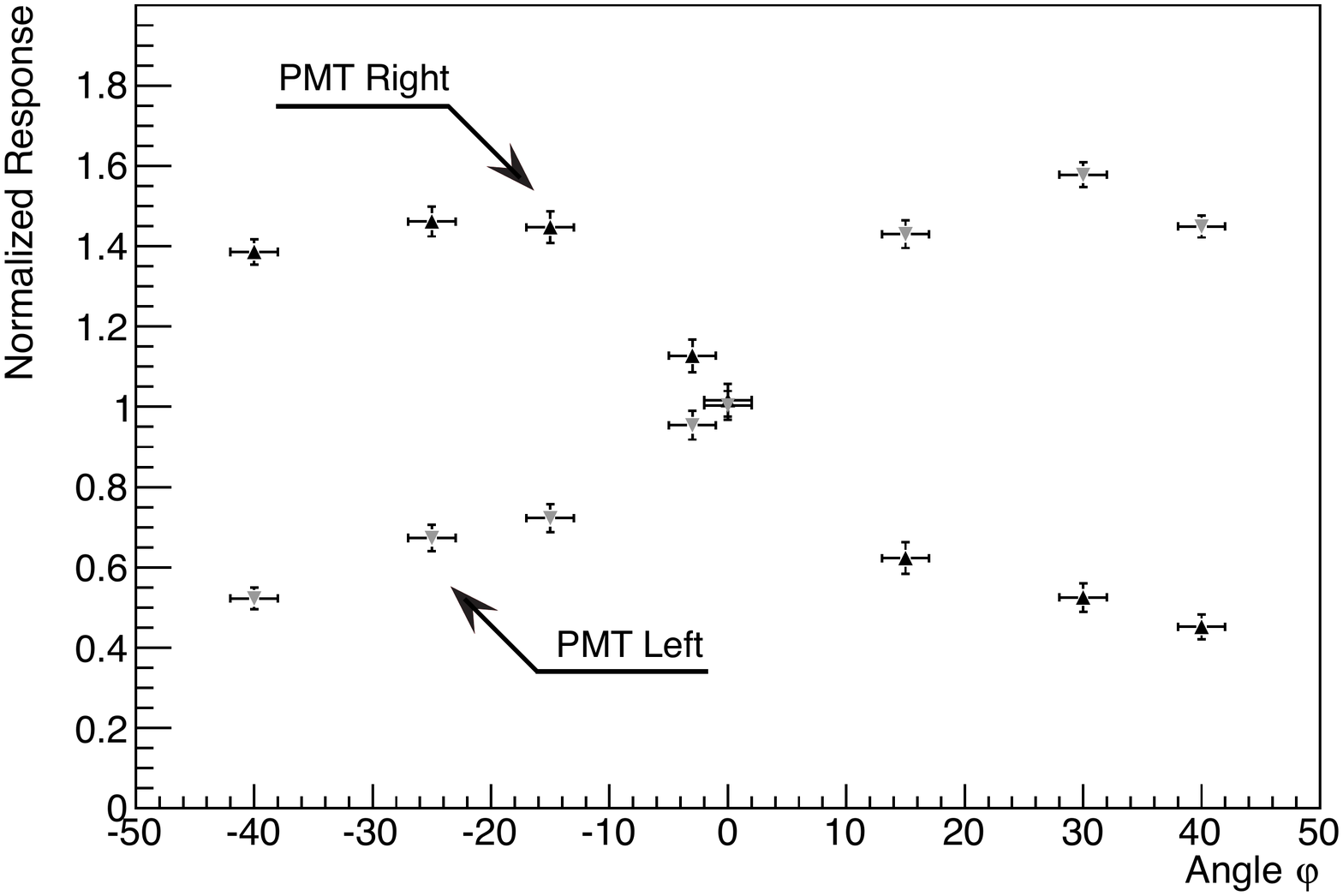}
\caption{Behavior of the responses corrected for the muon path length.}
\label{fig:charge-eq}
\end{centering}
\end{figure}

The dependence of $L(\varphi)$ and $R(\varphi)$ on
the angle $\varphi$ are shown in Fig.~\ref{fig:charge-eq}.
The two sides show the same behavior. Let's analyze the case of PMT-Left.
The detected light, corrected for the path length of the muon within the crystal,
is small and weakly dependent on the angle for $\varphi$ far from $\varphi_m$.
It shows a marked increase as long as
$\varphi$ approaches the value of $\varphi_m$ where 
the transmission of the \v{C}erenkov light 
is expected to have a maximum. 
At angles much larger than $\varphi_m$ a decrease of the amount
of light reaching PMT-Left is also visible. A symmetric analysis applies to PMT-Right.
This dependence on $\varphi$ of the signals on the two sides of the crystal
is a clear indication that a good fraction
of the light is due to \v Cerenkov photons.

\begin{figure}[h]
\begin{centering}
\includegraphics[width=10cm]{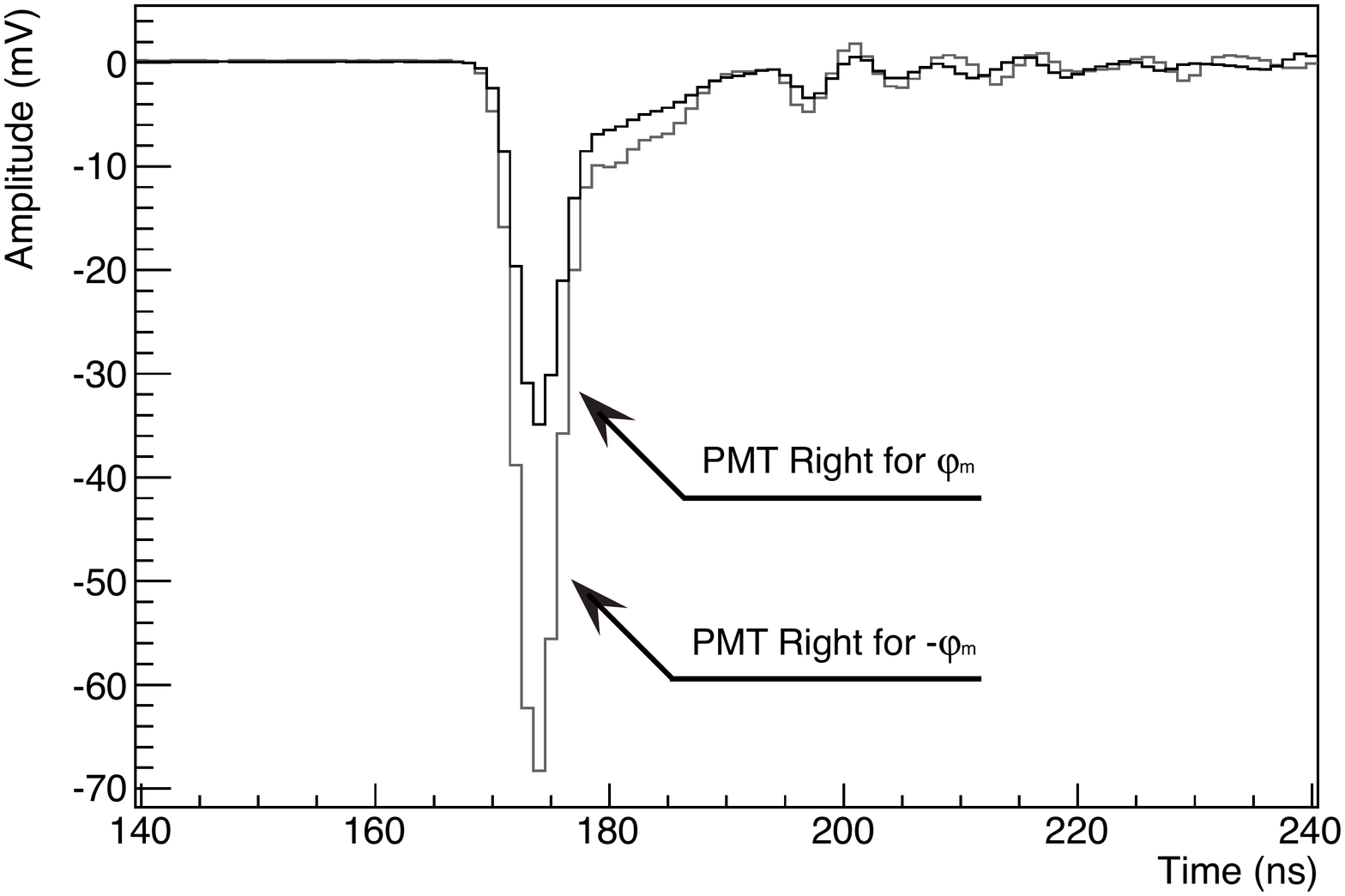}
\caption{Comparison between the average waveforms of the signals provided by 
PMT-Right as obtained for $\varphi = \varphi_m$  and $\varphi = -\varphi_m$.}
\label{fig:sig_17_24}
\end{centering}
\end{figure}

In order to understand the nature of the flat component,
the average waveforms of PMT-Right 
obtained for $\varphi = \varphi_m$ and $\varphi = -\varphi_m$
are reported in Fig.~\ref{fig:sig_17_24}.

Although the amplitudes are different, the signal shapes are the 
same. In particular, even for $\varphi = \varphi_m$, where  
\v Cerenkov photons cannot directly reach PMT-Right,
the signal is very fast and does not show any slow or exponential tail.
This indicates that also the flat component is likely due
to \v Cerenkov light able to reach the PMTs by
means of internal diffusion.

\section{The charge asymmetry}

\begin{figure} 
\begin{centering}
\includegraphics[width=10cm]{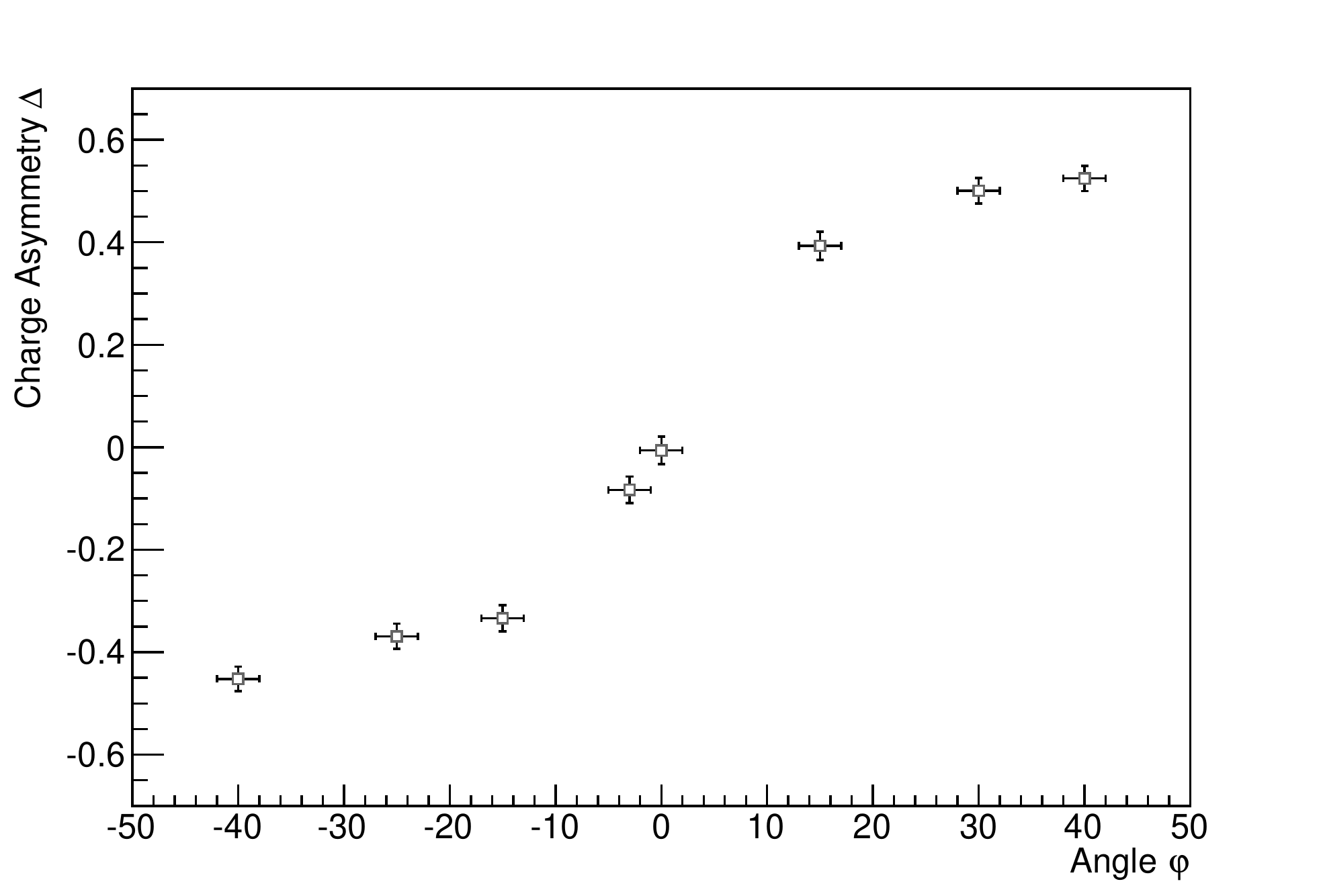}
\caption{Behavior of the charge asymmetry $\Delta$ as a function of the angle $\varphi$.}
\label{fig:delta}
\end{centering}
\end{figure}

In order to evaluate the ratio between the isotropic component of the light 
yield and the one that depends on $\varphi$, the {\it charge asymmetry}
$\Delta (\varphi)$ is studied. It is defined as:
\begin{equation}
\label{eq:delta}
\Delta(\varphi) = \frac{L(\varphi)-R(\varphi)}{L(\varphi)+R(\varphi)}=
\frac{B(\varphi)-B(-\varphi)}{2A+B(\varphi)+B(-\varphi)}
\end{equation}
and its behavior is shown in Fig.~\ref{fig:delta}.
For $\varphi \simeq \varphi_m$ the angle-dependent component of the 
light reaches its maximum ($B_{max}$) and, on the other hand,
$B(-\varphi_m)~=~0$. Therefore:
\begin{equation}
\label{eq:delta2}
\Delta(\pm \varphi_m) = \pm \frac{B_{max}}{2A+B_{max}}
\end{equation}
From the analysis of the data shown in Fig.~\ref{fig:delta} it results that 
$\Delta (-\varphi_m) \simeq -$~0.45 and $\Delta (\varphi_m) \simeq$ 0.55 
that means two values for $A$: 0.41 $B_{max}$ and 0.61 $B_{max}$.
According to the average of our measurements, the ratio between the 
component of the detected light that depends on the angle $\varphi$ 
and the total one is 0.66. 
This value represents a lower limit of the
\v Cerenkov component value that, thus, results
to be at least the $66\%$ of the total light yield.

\section{Conclusion}
The performed measurements show that a TeO$_2$ crystal emits light when crossed
by a charged particle.
The signals are very fast, having a rise time and decay time of the order of few nanoseconds.
The amount of light exiting from the crystal has a clear 
dependence on the angle $\varphi$
between the particle and the crystal.
The maximum of the light is collected 
for a value of $\varphi$ compatible with the one expected to maximize \v Cerenkov 
light output ($\varphi_m$).
A three times smaller amount of light is also detected for angles far from $\varphi_m$.
Most likely this is \v Cerenkov light diffused by the crystal lateral faces.
However, the measurements reported in the present paper allow to conclude 
that \v Cerenkov light represents at least the 66\% of
all the light emitted by a TeO$_2$ crystal.

\end{document}